\documentclass[prb,twocolumn,amssymb] {revtex4}
\usepackage{graphicx}

\begin{document}

\title{Two Generalizations of $\eta$ Pairing in Extended Hubbard Models}
\author{Hui Zhai
\\\it{Center for Advanced Study, Tsinghua University, Beijing,
100084, China} }
\date{\today}
\begin{abstract}
The Hubbard model in bipartite lattice has been rigorously proved
to have $\eta$-paired state as its eigenstates in Ref.\cite{Yang}.
In this paper, this result is first generalized to triangular
lattice, and then even arbitrary lattice structure. Then we also
consider its generalization to higher-spin Hubbard models. The
conditions under which these models have $\eta$-paired states as
its eigenstates were obtained.
\end{abstract}
\maketitle

\section{Introduction}

The fermionic model can possess off-diagonal long range
order(ODLRO) and thus becomes superconducting via particle
pairing\cite{YangRMP}. The usual BCS state does have ODLRO via the
mechanism of Cooper pairing, but it is not an eigenstate of a
Hamiltonian with a local interacting term. Thus, it is of
long-standing interests to find out models whose eigenstates
possess ODLRO. One of the exact results about this has been
obtained in Ref.\cite{Yang}, in which a local two-particle pairing
operator called $\eta$-pairing was constructed and proved to be an
eigen-operator of spin-$1/2$ Hubbard model in square lattice. Thus
the states created by the $\eta$ operator are exact eigenstates of
the Hubbard model and display ODLRO.

As one of the few rigorous results about the Hubbard model in
higher dimension, the $\eta$ pairing has attracted a lot of
theoretical interests since it was proposed. Although in the
strong repulsive interaction case these eigenstates created by the
$\eta$ operator, which are all doubly-occupied states, have a
large excitation energy, and therefore is not closely related to
current discussion of high-Tc superconductivity of Cuprates in the
content of $t-J$ model, the $\eta$ operator has been found very
helpful and widely used in the study of attractive Hubbard
model\cite{Singh}\cite{Betsuyaku}\cite{Stein}. Although in the
majority of cases the $\eta$-pairing state is an excited state, it
has been found to be ground state for some extended Hubbard
models, such as in presence of nearest-neighbor interaction,
bond-charge interaction and so on\cite{Boer}\cite{Campbell}.
Besides, the properties of $\eta$-pairing operator have also been
used to construct an $SU(2)$ symmetry group of the Hubbard
model\cite{Zhang}. Together with another $SU(2)$ group generated
by particle-hole operators, an $SO(4)$ symmetry of the Hubbard
model has been found, which unifies the spin order and
superconductor pairing order\cite{Yangzhang}.

Recently, rapid theoretical and experimental development has been
made in the field of ultracold atoms, among which the successful
application of optical lattice technique is a remarkable
one.\cite{BlochMott} The fermionic atoms confined in optical
lattice can be described by (extended) Hubbard models. Using the
technique of quantum optics, various lattice structures can be
realized\cite{KagomeLattice}. It was also proposed that the
tunnelling elements between nearest lattice sites can also be
experimentally controlled so that the atoms in optical lattice can
be effectively viewed as charged particles moving in a magnetic
field\cite{Lukin}\cite{Muller}. In this paper, we will try to
generalize the $\eta$-paring to the triangular lattice, and even
to arbitrary lattice structures.

On the other hand, the lattice atomic gases also provide a testing
grounds to study high-spin extension of Hubbard models, and the
interaction between atoms is spin-dependent and can be tuned in a
wide range due to the Feshbach resonance technique. Recently, Wu,
Hu and Zhang discussed interacting spin-$3/2$ fermion atoms in
optical lattice. They found that this model not only possesses a
generic $SO(5)$ symmetry formed by particle-hole operators, but
also has an $SU(2)$ symmetry constructed by $\eta$ paring
operators in the condition of some special interaction
parameters.\cite{SCZhang} We will also discuss the $\eta$-pairing
for more higher spin models in the following.

We organize this report as follows. In the second section we will
briefly review the keys to obtain the commutation relation
$[\hat{\eta}^\dag, \hat{H}]=C\eta^\dag$ for spin-$1/2$ Hubbard
model in square lattice, where $\hat{H}$ is the Hamiltonian and
$C$ is a constant. In the third section we will generalize it to
Hubbard model in triangular lattice. Because the triangular
lattice is not a bipartite lattice, the $\eta$ operator is
generally not an eigen-operator of the kinetic energy term.
However, we find for a class of special hopping elements, a
modified $\eta$ operator can be eigen-operator. In the fourth
section, we will discuss the interacting fermions with higher
spins. We will answer the question that how to choose the
interacting parameters so that an extended $\eta$ operator can be
an eigen-operator.

\section{The $\eta$ Pairing}

Before discussing the generalization, we should first remind
ourselves of some basic ideas of the $\eta$ pairing discussed in
Ref.\cite{Yang} for the Hubbard model on two-dimensional square
lattice, whose Hamiltonian is written as
\begin{equation}
\hat{H}=\hat{T}+\hat{V}=-t\sum\limits_{ij}(a_{\uparrow i}^\dag
a_{\uparrow j}+a_{\downarrow i}^\dag a_{\downarrow
j}+h.c.)+U\sum\limits_{i}n_{\uparrow i}n_{\downarrow
i}.\label{Hubbard}
\end{equation}
A two-particle pairing operator of spin-$1/2$ fermions can be
generally written as
\begin{equation}
\eta=\sum\limits_{\vec{k}}g(\vec{k})a_{\uparrow \vec{k}}^\dag
a_{\downarrow \vec{q}-\vec{k}}^\dag,
\end{equation}
it is an eigen-operator of the interaction energy term when
$\langle n^\prime| \hat{V}|n^\prime\rangle-\langle
n|\hat{V}|n\rangle$ is a constant for any $|n\rangle$ and
$|n^\prime\rangle$, for which $\langle
n|\eta^\dag|n^\prime\rangle$ is nonvanishing. Here the states
$|n\rangle$ are all eigenstates of the interaction term and they
form a complete bases of the Hilbert space. Noticing that the
interaction terms are all local, the pairing should also be a
local one, which means that $g(\vec{k})$ should be taken as a
constant independent of $\vec{k}$.

In the momentum space the kinetic energy term can be written as
\begin{equation}
\hat{T}=\sum\limits_{\vec{k}\sigma}f(\vec{k})a_{\vec{k},\sigma}^\dag
a_{\vec{k},\sigma}
\end{equation}
with
\begin{equation}
f(\vec{k})=-2t(\cos k_{x}+\cos k_{y}).
\end{equation}
In Ref.\cite{Yang} it is noticed that the $\eta$ operator commutes
with the kinetic energy term when $\vec{q}$ is taken as
$\vec{\pi}$ because
\begin{equation}
\cos k +\cos (\pi-k)=0.
\end{equation}
Thus in the coordinate space the $\eta$ pairing is a superposition
of local particle pairs
\begin{equation}
\eta=\sum\limits_{\vec{r}}e^{i\vec{\pi}\cdot\vec{r}}a_{\uparrow
\vec{r}}^\dag a_{\downarrow \vec{r}}^\dag,
\end{equation}
where one essential point is the sign of the pair in one site
should be different from those in its nearest-neighbors. Hence the
$\eta$ operator thus defined satisfies the commutation relation
\begin{equation}
\eta^\dag H-H\eta^\dag=-2U\eta^\dag.
\end{equation}

\section{Generalization to Triangular Lattice}

We notice that the above discussion about $\eta$ pairing relies on
whether the lattice structure is bipartite, because the lattice
structure directly determines the explicit form of $f(\vec{k})$,
which plays a crucial role in finding $\eta$ operator. Hence, for
the Hubbard model in triangular lattice, the $\eta$ paring
operator introduced in previous section is not generally an
eigen-operator. Then the question arouses that under which
condition the model has a pairing operator as its eigen-operator.

The Hamiltonian of the Hubbard model in a triangular lattice under
consideration is written as
\begin{equation}
\hat{H}=\sum\limits_{ij\sigma}t_{ij}a_{i\sigma}^\dag
a_{j\sigma}+U\sum\limits_{i}n_{\uparrow i}n_{\downarrow i}
\end{equation}
First focusing on the case with translational invariance, the
kinetic energy term $\hat{T}$ can be written in a general form
\begin{eqnarray}
\hat{T}=-t\sum\limits_{i\sigma}\left(e^{i\theta_{1}}a_{ij,\sigma}^{\dag}a_{i+1j,\sigma}
+e^{i\theta_{2}}a_{ij,\sigma}^\dag a_{i+\frac{1}{2}
j+\frac{\sqrt{3}}{2},\sigma}\right.\nonumber\\
\left.+e^{i\theta_{3}}a_{ij}^\dag a_{i+\frac{1}{2}
j-\frac{\sqrt{3}}{2},\sigma}+h. c.\right).
\end{eqnarray}
Here $\theta_{i}$ are the phases defined in links, whereas only
$\prod_{\text{closed loop}}e^{i\theta_{i}}$ are gauge invariant
quantities, which represent the flux number through the closed
loops. In the momentum space, the kinetic energy term can be
written as
%\begin{widetext}
\begin{eqnarray}
&&\hat{T}=-2t\sum\limits_{k_{x}k_{y}\sigma}\left[\cos\left(k_{x}+\theta_{1}\right)+
\cos\left(\frac{k_{x}}{2}+\frac{\sqrt{3}}{2}k_{y}+\theta_{2}\right)\right.\nonumber\\
&&\left.+\cos\left(\frac{k_{x}}{2}-\frac{\sqrt{3}}{2}k_{y}+\theta_{3}\right)\right]a_{\vec{k},\sigma}^\dag
a_{\vec{k},\sigma}\nonumber\\
&&=-4t\sum\limits_{k_{x} k_{y}\sigma}
\left[\cos\left(\frac{k_{x}}{2}+\frac{\theta_{1}}{2}-\frac{\pi}{4}\right)
\sin\left(\frac{\pi}{4}-\frac{\theta_{1}}{2}-\frac{k_{x}}{2}\right)\right.\nonumber\\
&&\left.+
\cos\left(\frac{k_{x}}{2}+\frac{\theta_{2}+\theta_{3}}{2}\right)
\cos\left(\frac{\sqrt{3}}{2}k_{y}+\frac{\theta_{2}-\theta_{3}}{2}\right)\right]a_{\vec{k}\sigma}^\dag
a_{\vec{k}\sigma}.\nonumber\\
\end{eqnarray}
%\end{widetext}

We notice that when $\theta_{1}-\theta_{2}-\theta_{3}=\pi/2$, the
expression of $f(\vec{k})$ turns out to be
\begin{eqnarray}
&&f(\vec{k})=-4t\cos\left(\frac{k_{x}}{2}+\frac{\theta_{1}}{2}-\frac{\pi}{4}
\right)\nonumber\\
&&\left[\sin\left(\frac{\pi}{4}-\frac{\theta_{1}}{2}-\frac{k_{x}}{2}\right)+
\cos\left(\frac{\sqrt{3}}{2}k_{y}+\frac{\theta_{2}-\theta_{3}}{2}\right)
\right],\nonumber\\
\end{eqnarray}
In this case, because
\begin{equation}
f(k_{x},k_{y})+f\left(2\pi-k_{x},\frac{2}{\sqrt{3}}(\theta_{3}-\theta_{2})-k_{y}\right)=0
\end{equation}
the $\eta$ operator defined as
$\eta^\dag=\sum\limits_{\vec{k}}a_{\uparrow \vec{k}}^\dag
a_{\downarrow \vec{q}-\vec{k}}^\dag$ with $\vec{q}$ taken as
$(2\pi, \frac{2}{\sqrt{3}}(\theta_{3}-\theta_{2}))$ is an
eigen-operator of this model.

\begin{figure}[htbp]
\begin{center}
\includegraphics[width=2.3in]
{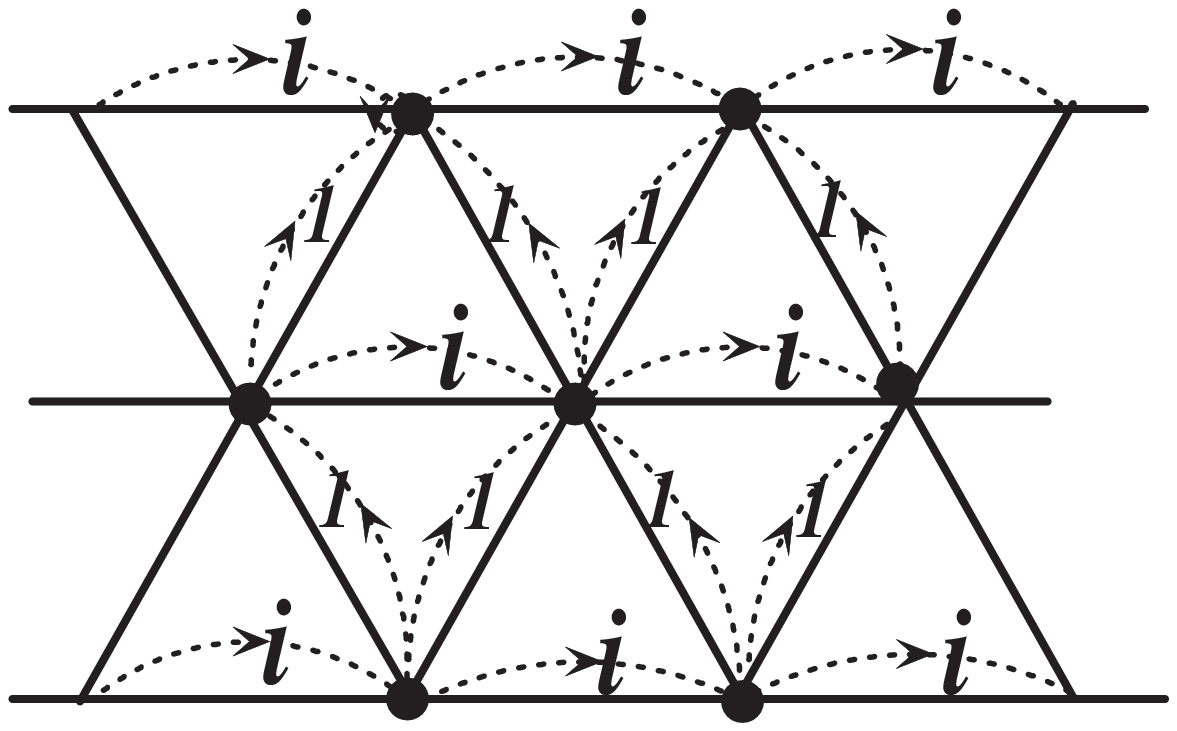}
\includegraphics[width=2.4in]
{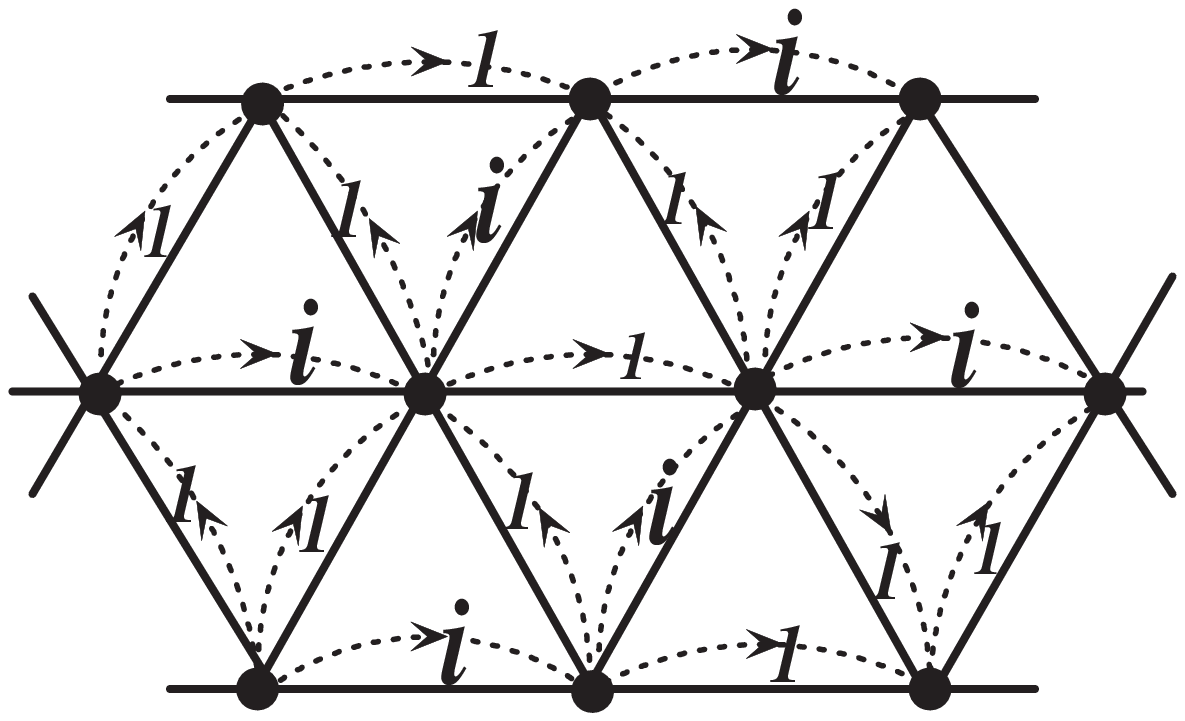} \caption{Two typical configurations of hopping
elements $t_{ij}$, which satisfy the requirement
Eq.(\ref{criterion}). The number $1$ or $i$ denotes the hopping
elements $t_{ij}$ form one site to another site. For the upper
one, the Hamiltonian has translational invariance, and for the
lower one it is not translational
invariant.\label{trangluarlattice}}
\end{center}
\end{figure}

Therefore we have succeeded in obtaining a sufficient condition
for the $\eta$ pairing states being eigenstates, which is
$\theta_{1}-\theta_{2}-\theta_{3}=\pi/2$. A typical configuration
is plotted in Fig.\ref{trangluarlattice}. This condition in fact
means that there is a $\pi/2$ flux through each plaquette.

Furthermore, we notice that the $\eta$ operator is a local
operator, and the kinetic energy term only involves
nearest-neighbors hopping, the assumption of translational
invariance is indeed not essential. Let
$t_{ij}=|t|e^{i\theta_{ij}}$, the condition can be further
extended to
\begin{equation}
\sum_{C}\theta_{ij}=\frac{\pi}{2}(\text{mod}
\pi),\label{criterion}
\end{equation}
where $C$ stands for any closed loop with odd numbers of links.
This criterion can be easily verified in the coordinate space
representation, and it is valid for any other lattice structure.
One can see that the bipartite lattice is just a special case of
this criterion, because any closed loop in a bipartite lattice
must contain even numbers of links.

\section{Generalization to High Spins Fermions}

In this section we will consider an interacting
spin-$\frac{2n-1}{2}$ fermions model, with $2n$ different spin
states in each site. In this model, the two-particle pairing
operator is generally not eigen-operator of the interaction term
because of the enlargement of the Hilbert space. It is our
interest to find out an extended $\eta$ operator and when the
pairing operator is an eigen-operator of the interaction term
$\hat{V}$.

We consider the extended Hubbard model as
\begin{equation}
\hat{H}=\sum\limits_{ij\sigma}t_{ij}a^\dag_{i\sigma}a_{j\sigma}+\hat{V},
\end{equation}
where $t_{ij}$ satisfy the criterion Eq.(\ref{criterion}). It can
be shown that a better choice of the on-site interaction should be
like
%\begin{widetext}
\begin{eqnarray}
&&\hat{V}=V_{2}(a_\uparrow^\dag a_\uparrow a_\downarrow^\dag
a_\downarrow+b_\uparrow^\dag b_\uparrow b_\downarrow^\dag
b_\downarrow+c_\uparrow^\dag c_\uparrow c_\downarrow^\dag
c_\downarrow+\cdot\cdot\cdot)\nonumber\\&&
+V_{0}\sum\limits_{\sigma,\sigma^\prime}(a_\sigma^\dag a_\sigma
b_{\sigma^\prime}^\dag b_{\sigma^\prime}+a_\sigma^\dag a_\sigma
c_{\sigma^\prime}^\dag c_{\sigma^\prime}+b_{\sigma}^\dag
b_{\sigma}c_{\sigma^\prime}^\dag
c_{\sigma^\prime}+\cdot\cdot\cdot)\nonumber\\&&+V_{1}(a_{\uparrow}^\dag
a_{\downarrow}^\dag
b_{\uparrow}b_{\downarrow}+h.c.+a_{\uparrow}^\dag
a_{\downarrow}^\dag
c_{\uparrow}c_{\downarrow}+h.c.+\cdot\cdot\cdot).
\label{interacting}
\end{eqnarray}
%\end{widetext}
Here we denote $a_\uparrow$ and $a_{\downarrow}$ as fermionic
annihilation operators for $\pm 1/2$ states, $b_{\uparrow}$ and
$b_\downarrow$ for $\pm 3/2$ states and so on, i.e.
\begin{equation}
[a_{\sigma},
a^\dag_{\sigma^\prime}]_{+}=\delta_{\sigma\sigma^\prime}
\end{equation}
and
\begin{equation}
[a_{\sigma}, b^\dag_{\sigma^\prime}]_{+}=0.
\end{equation}

The first term describes the interaction between $m/2$ state and
$-m/2$ state, the second describes the interaction between $\pm
m/2$ states and $\pm m^\prime/2$ states with different $m$ and
$m^\prime$, and the last term describes the spin flipping
processes in the channel of total $S_{z}$ equalling to zero. The
interacting between two spin-$3/2$ atoms discussed in
Ref.\cite{SCZhang} is a special case of this form. One can find
that the presence of the spin-flipping term is a consequence of
spin-dependent interaction.

It is illuminating to consider the commutation relation between
$a^\dag_{\uparrow} a^\dag_{\downarrow}$ and these terms. For the
first term, the commutation relation is
\begin{equation}
[a_\uparrow^\dag a_\downarrow^\dag, a_\uparrow^\dag a_\uparrow
a_\downarrow^\dag a_\downarrow]_{-}=-a_\uparrow^\dag
a_\downarrow^\dag.
\end{equation}
For the second term we have
\begin{equation}
[a_\uparrow^\dag a_\downarrow^\dag, a_\uparrow^\dag a_\uparrow
b_\sigma^\dag b_\sigma]_{-}=-a_{\uparrow}^\dag a_{\downarrow}^\dag
b_\sigma^\dag b_\sigma,
\end{equation}
and for the spin flipping term
\begin{equation}
[a_\uparrow^\dag a_\downarrow^\dag, b_{\uparrow}^\dag
b_{\downarrow}^\dag
a_{\uparrow}a_{\downarrow}]_{-}=-b_{\uparrow}^\dag
b_{\downarrow}^\dag a_{\uparrow}^\dag
a_{\uparrow}-b_{\uparrow}^\dag b_{\downarrow}^\dag
a_{\downarrow}^\dag a_{\downarrow}+b_{\uparrow}^\dag
b_{\downarrow}^\dag.
\end{equation}
Hence, to satisfy the requirement $\eta^\dag V-V\eta^\dag\propto
\eta^\dag$, the terms such as $a_{\uparrow}^\dag
a_{\downarrow}^\dag b_\sigma^\dag b_\sigma$ should be cancelled
with each other, which means that the $\eta$ operator should be
taken as
\begin{equation}
\eta=\sum\limits_{\vec{r}}e^{i\vec{q}\vec{r}}(a_{\vec{r}\uparrow}^\dag
a_{\vec{r}\downarrow}^\dag + b_{\vec{r}\uparrow}^\dag
b_{\vec{r}\downarrow}^\dag
+c^\dag_{\vec{r}\uparrow}c_{\vec{r}\downarrow}^\dag+\cdot\cdot\cdot),
\end{equation}
and the parameters in Eq.(\ref{interacting}) should satisfy the
requirement
\begin{equation}
V_{1}=-2V_{0}.\label{condition}
\end{equation}
This relation recovers the result obtained in Ref.\cite{SCZhang}.
The commutation relation is therefore
\begin{equation}
[\eta^\dag, V]=-(V_{2}+2V_{0})\eta^\dag,
\end{equation}
The generalized $\eta$-operator is also a superposition of local
spin-singlet two-particle pairing. The state created by the
operator has ODLRO for the same reason discussed in
Ref.\cite{Yang}and Ref.\cite{YangRMP}

\section{Conclusion and Discussion}

In summary, we have discussed two generalizations of the
$\eta$-paring to two kinds of extended Hubbard models. The
original conclusions about the $\eta$-pairing was restricted for
bipartite lattice structure, in this paper we showed that for the
Hubbard model defined on any lattice structure, a modified
$\eta$-operator can be the eigen-operator if the hopping elements
satisfying the criterion Eq.(\ref{criterion}). We also discussed
the $\eta$-pairing for higher-spin Hubbard models. We find that in
order that the model has modified $\eta$-paired states as its
eigenstates, the interaction part should contain spin-flipping
term and the parameters should satisfy the condition
Eq.(\ref{condition}).

As the original $\eta$-paired state\cite{Yang}, the generalized
$\eta$-paired states are also usually excited states. One thing of
much concern is when these generalized $\eta$-paired states become
ground states. In the original paper\cite{Yang}, it is shown the
$\eta$-paired state is metastable for attractive interaction. And
later, various extended models which have the $\eta$-paired state
as their ground states were proposed\cite{Boer}\cite{Campbell}.
Following these ideas, how to obtain models having the generalized
$\eta$-paired states as the ground states is under further
consideration. We hope that these efforts could shed light on the
study of fermionic atoms in optical lattice and some solid-state
materials such as $NaCoO$ which have triangular lattice
structure\cite{NocoO}.

{\it{Acknowledgement}:} The author would like to acknowledge
Professor C. N. Yang for his encouragement and guidance, and for
his suggestion to study this problem and spending a lot of time on
discussion sections. The author would also like to thank Professor
L. Chang, Z. Y. Weng, J. P. Hu, R. L$\ddot{u}$ and Dr. X. L. Qi,
Q. Zhou for helpful discussions. This work is supported by
National Natural Science Foundation of China.

\end{document}